\documentclass{sig-alternate-2013}
\usepackage{url}
\usepackage{graphicx}
\usepackage{epstopdf}
\usepackage{amsmath}
\usepackage{amssymb}
\usepackage{mathtools}
\usepackage{multirow}
\usepackage{colortbl}
\usepackage{xcolor}
\usepackage{subfigure}
\usepackage[font=small,format=plain,labelfont=bf,up,textfont=bf,up,justification=justified,singlelinecheck=false]{caption}
\DeclareMathOperator*{\argmin}{arg\,min}

\newtheorem{definition}{Definition}
\usepackage{booktabs}

\newfont{\mycrnotice}{ptmr8t at 7pt}
\newfont{\myconfname}{ptmri8t at 7pt}
\let\crnotice\mycrnotice%
\let\confname\myconfname%

\permission{Permission to make digital or hard copies of all or part of this work for personal or classroom use is granted without fee provided that copies are not made or distributed for profit or commercial advantage and that copies bear this notice and the full citation on the first page. Copyrights for components of this work owned by others than ACM must be honored. Abstracting with credit is permitted. To copy otherwise, or republish, to post on servers or to redistribute to lists, requires prior specific permission and/or a fee. Request permissions from permissions@acm.org.}
\conferenceinfo{CIKM'15,}{October 19-23, 2015, Melbourne, VIC, Australia.}
\copyrightetc{Copyright 2015 ACM \the\acmcopyr}
\crdata{978-1-4503-3794-6/15/10\ ...\$15.00\\
http://dx.doi.org/10.1145/2806416.2806486}

\begin{document}
\newcommand{\tuan}[1]{{\textcolor{blue}{[\textit{\small Tuan: #1}]}}}

\newcommand\vpar{{\vspace*{1em}}}
\newcommand{\para}[1]{\noindent{\textbf{#1.}}}
\newcommand{\parai}[1]{\noindent{\textit{#1.}}}
\newcommand{\term}[1]{{\it {\small #1}}}

%

\title{Balancing Novelty and Salience:\\
Adaptive Learning to Rank Entities for Timeline\\
Summarization of High-impact Events}
%
\numberofauthors{1}
\author{
\alignauthor
Tuan Tran, Claudia Nieder\'{e}e, Nattiya Kanhabua, Ujwal Gadiraju, and Avishek Anand\\
\affaddr{$ $}\\
       \affaddr{L3S Research Center / Leibniz Universit\"{a}t Hannover, Germany}\\
\vspace{5pt}
       \email{\{ttran, niederee, kanhabua, gadiraju, anand\}@L3S.de}
}

\maketitle
\begin{abstract}
Long-running, high-impact events such as the Boston Marathon bombing often develop through many stages and involve a large number of entities 
in their unfolding.
Timeline summarization of an event by key sentences eases story digestion, but does not distinguish between what a user remembers  and what she might want to re-check. In this work, 
we present a novel approach for timeline summarization of high-impact events, which uses entities instead of sentences 
for summarizing the event at each individual point in time.  Such entity summaries can serve as both  
(1)~important memory cues in a retrospective event consideration and (2)~pointers for personalized event exploration. 
In order to automatically create such summaries, it is crucial to identify the ``right" entities for inclusion. 
We propose to learn a ranking function for entities, with a dynamically adapted trade-off between the in-document salience of entities
and the informativeness of entities across documents, i.e., the level of new information associated with an entity for a
time point under consideration. 
Furthermore, for capturing collective attention for an entity we use an innovative soft labeling approach based on Wikipedia.
Our experiments on a real large news datasets confirm the effectiveness of the proposed methods.
\end{abstract}

\vspace{0.2cm}
{\small
\noindent {\bf Categories and Subject Descriptors} H.3.3 [Information Storage and Retrieval]: Information Search and Retrieval\\
\noindent {\bf General Terms} Algorithms, Experimentation}


\section{Introduction}
\label{sec:intro}

\begin{figure}
\centering
\includegraphics[width=\columnwidth]{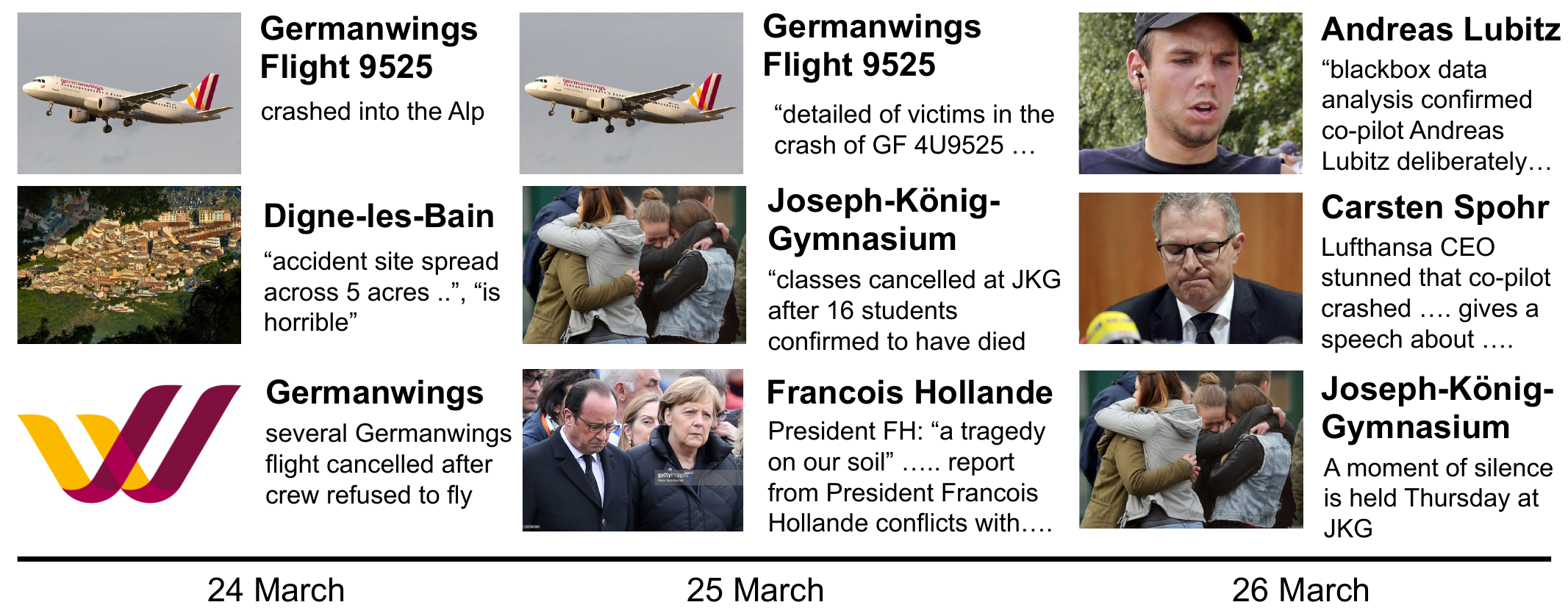}
\caption{Entity Timeline for the 2015 Germanwings Plane Crash Event}
\label{fig:4u9525}
\end{figure}

High-impact, real-world events often unfold in an unexpected way, dynamically involving a variety of entities. 
This is especially true for long-running events
when full information about the event and its development becomes available 
only in the course of days after the happening, 
as in the case of a recent Germanwings airplane crash. 
In this paper, we study event timeline summarization and present a novel method 
which shows key entities at different time points of an
event thus capturing this dynamic event unfolding. 
In contrast to other work in event summarization \cite{yan2011evolutionary,McCreadie:2014:IUS:2661829.2661951}, our \textit{entity timelines} use entities instead of sentences as main units of summarization
as depicted in the case of the 2015 Germanwings plan crash (Figure \ref{fig:4u9525}).
Such summaries can be easily digested and used both as starting points for personalized exploration of event details, and for retrospective 
revisiting. 
The latter can be triggered by 
a new similar event,
or by a new twist in the story. For example, 
the testimonial of the captain
in the Costa Concordia trial in late 2014
triggered a revisiting of the 
disaster in 2012.

From a cognitive perspective, for event revisiting, we rather create ``memory cues" 
to support remembering the unfolded events than summaries for rehearsing the details. In fact, memory cues can be regarded as ``a circumstance or piece of information which aids the memory in retrieving details not re-called spontaneously" (Oxford online dictionary, 2015). In this sense, our work is related to the idea of designing or creating memory cues for real-life remembering~\cite{HoEg14}. 
Entities such as 
persons and locations have been identified as very effective external memory cues~\cite{berntsen2009involuntary}.
In addition, the importance of entities in event summarization has also been shown in recent 
work~\cite{meng2012entity,moshfeghi2013influence}.



For creating an entity timeline,
the entities to be used in the summary have to be chosen carefully. 
They should 1)~be characteristic for a respective time point 
of the event, 2)~not be repetitive (if nothing new happened 
with respect to the entities), 3)~be associated to relevant event information,
and 4)~be interesting to the reader. For this purpose, 
we propose an approach to entity summarization, which dynamically 
combines \textit{entity salience} with the \textit{informativeness} 
of entities at a considered point in time.
Entity salience, on the one hand, considers the property of being 
in the focus of attention in a document has been studied in previous work~\cite{boguraev1999salience,gamon2013identifying,dunietz2014new}.
In~\cite{boguraev1999salience}, Boguraev and Kennedy use salient text phrases for the creation of so-called \textit{capsule overviews}, whereas recently methods for the identification of salient entities, e.g., in Web pages~\cite{gamon2013identifying} and news articles 
\cite{dunietz2014new}, have been developed. Informativeness, on the other hand, assesses
the level of new information associated with an entity in a text and can be 
computationally measured using features derived from statistical and linguistic information~\cite{wu2013measuring}.

In more detail,
we aim at optimizing a trade-off between 
the in-document salience of entities and the informativeness 
of entities across documents describing the unfolding of an event. 
Our contributions are:

\vspace{-0.1cm}
\begin{itemize}
    \item We are the first to propose an entity timeline for novel user experience in event exploration and digestion.
\vspace{-0.1cm}
    \item We propose a novel entity ranking model that dynamically learns a trade-off between entity salience and informativeness.  
\vspace{-0.1cm}
    \item For training our ranking model, we propose a method for gathering soft labels (or ground truths) by mining collective attention in Wikipedia.
\end{itemize}
\vspace{-0.1cm}

We validate our contributions with experiments on a large real-world news dataset covering various types of events. The rest of the chapter is organized as follows. Section 2 discusses the related work. Sections 3,4 details our proposed method. Section 5 discusses features used in learning framework. Section 6 presents the experiments and results. Finally, section 7 concludes our paper.

\section{Related Work}

\para{News Summarization}
The task of news summarization has been already studied in various contexts, which range from focusing on multi-document summarization~\cite{boguraev1999salience,erkan2004lexrank} to generating a timeline summary for a specific news story~\cite{yan2011evolutionary,McCreadie:2014:IUS:2661829.2661951,tran2013leverage,zhao2013timeline}. News stories can be complex, having a non-linear structure and associated to multiple aspects. Shahaf et al.~\cite{shahaf2012trains} propose a novel method for summarizing complex stories using metro maps that explicitly capture the relations among different aspects and display a temporal development of the stories. Instead of using documents or sentences as a information units, we provide a set of entities, as \textit{memory cues}, for supporting event exploration and digestion at each individual point in time. 
To the best of our knowledge, none of the previous work 
provides a trade-off solution that balances between content-based and collective attention-based approaches, in supporting the entity-centric summarization.

\para{Entity Ranking and Entity Salience}
In recent years, entity retrieval and ranking have gained interest from the IR community moving beyond traditional document retrieval to more specific information about entities in response to a query, e.g.,
~\cite{meng2012entity,moshfeghi2013influence}. 
Although a lot of interesting work on entity ranking has already been proposed, most of previous works have focused on static collections, thus ignoring the temporal dynamics of queries and documents. The relevant work closest to ours in this respect is by Demartini et al.~\cite{demartini2010taer}, where the task tackled is to identify the entities that best describe the documents for a given query. The entities are identified by analyzing the top-k  retrieved documents at the time when the query was issued as well as relevant documents in the past.
In IR, salient entities in news documents help improving the performance of document retrieval~\cite{moshfeghi2013influence}. There is a body of work in identifying salient entities in general Web~\cite{gamon2013identifying} and in news domain~\cite{dunietz2014new}, but the existing work does not take into account the time dimension. Our work targets a different question, using entities as a pivot, and it considers the global context (timeline of the event) as a new dimension of entity salience determination.

\para{Learning to Rank Multiple Criteria}
In our work, we apply a learning to rank (L2R) technique to generate timeline summaries in a similar way as done in~\cite{tran2013leverage,McCreadie:2014:IUS:2661829.2661951}. In contrast to the previous work, we propose a novel joint learning method, which optimizes different aspects of an event. The model is able to distinguish the aspects semantically. We also incorporate the time effect (i.e., decay) into the joint model, making our learning to rank time-sensitive and suitable for timeline summarization.

\section{Overview}

\subsection{Preliminaries}

\para{Long-running Event} Following \cite{McCreadie:2014:IUS:2661829.2661951}, we define a long-running event as ``a newsworthy happening in the world that is significant enough to be reported on over multiple days''. 
Each event is represented by a short textual description or a set of keywords $q$, where we will use $q$ to denote the event in the rest of this paper. 
For example, the bombing incident during Boston Marathon in April 2013 can be described by the terms ``boston marathon bombing''. We assume that the relevant time frame is
split into a sequence of consecutive non-overlapping, equal-sized time intervals $T =\{t_1,\ldots, t_n\}$ in a chronological order, in our case individual days. Furthermore, for a given event $q$, there is a set of timestamped documents $D_q$ (each with a publication date) reporting on the event.  
We define a \textit{reporting timeline} $T_q =\{t_{k_1}, \ldots, t_{k_j}\}$ as 
an ordered list of (not necessarily consecutive) those time periods $t_{k_i}$ in $T$, which contain the publication date of at least document in $D_q$. Finally, we denote the set of all documents about $q$ published within a time period $t_{k_i}$ as $D_{q,i}$.


\vpar
\para{Entity} 
We are interested in named entities mentioned in documents, 
namely, persons, organizations, locations. An entity $e$ can be identified by a canonical name, and can have multiple terms or phrases associated with $e$, called \textbf{labels}, which refer to the entity. We call an appearance of an entity label in a document, a \textbf{mention} of $e$, denoted by $m$. We define
$E_q$ as the set of all entities mentioned in $D_q$.
Furthermore, we define the text snippet surrounding $m$ (e.g., sentence or phrases) as the \textbf{mention context}, denoted $c(m)$. 

\vpar
\para{Entity Salience and Informativeness} Similarly to~\cite{dunietz2014new}, we define the entity salience as the quality of ``being in the focus of attention'' in the corresponding documents. 
Another relevant aspect considered for selecting entities to be included in an event timeline is \emph{ informativeness}~\cite{gamon2013identifying}, which imposes that selected entities in an evolving event should also deliver novel information when compared to the past information. For example, although the airline "Germanwings" stays relevant for many articles reporting on the plane crash, it will only be considered as informative, if new information about the 
airline becomes available. 

\vpar
\para{Problem} 
Given a long-running event $q$, a time interval $t_i$ in its reporting timeline $T_q$, and the set of entities $E_q$, we aim to identify the top-$k$ salient and informative entities for supporting  the exploration and digestion of $q$ at $t_i$.

\subsection{Approach Overview}
We tackle this problem by learning a function for ranking entities, which is aimed at optimizing the trade-off between in-document entity salience and the informativeness of entities across documents:

\vspace*{-1em}
\begin{equation}
\label{eq:problem}
y^{(q)}_t = f(\mathbf{E}, \omega_{\mathrm{s}}, \omega_{\mathrm{i}}), f\in \mathcal{F}
\end{equation}

\noindent where $y^{(q)}_t$ is the vector of ranking scores for entities in $E_q$ at time interval $t$,
$\mathbf{E}$ is a matrix composed from feature vectors of entities in $E_q$ extracted from their mention contexts. $\omega_{\mathrm{s}}, \omega_{\mathrm{i}}$ are the unknown parameter vectors for ranking entities based on salience and informativeness, respectively. 

In our work, a ranking function is based on a learning-to-rank technique~\cite{joachims2002optimizing}.
A general approach for learning-to-rank is to optimize a defined loss function $L$ given 
manual annotation or judgments $\mathbf{y}^{(q)}_{j}$ of entities for a set of training events $\mathcal{Q}$ within the time intervals $T_q$:
\begin{equation}
\label{eq:l2r}
\hat{f} = \argmin_{f\in \mathcal{F}} \sum_{q\in \mathcal{Q}}\sum_{t_j\in T_q} L(f(\mathbf{E}^{(q)}_{j},\omega_{\mathrm{s}},\omega_{\mathrm{i}}),\mathbf{y}^{(q)}_{j}) 
\end{equation}

Two major challenges must be taken into account when learning a ranking function defined in Equation \ref{eq:l2r}. First, we need a reliable source for 
building judgments (ground truths) for annotating entities by considering their salience with respect to a given event.
In addition, the judgments must be dynamically adapted to the evolving of entities along the unfolding event, i.e., bearing of novel information. Second, the models of our two aspects  $\omega_{\mathrm{s}},\omega_{\mathrm{i}}$ must be unified to produce 
a joint learned function for ranking entities. In the subsequent sections, we will explain our proposed method for these challenges in more detail.

\subsection{Framework}
\label{sec:archi}

\begin{figure}
\centering
\includegraphics[width=1.0\columnwidth]{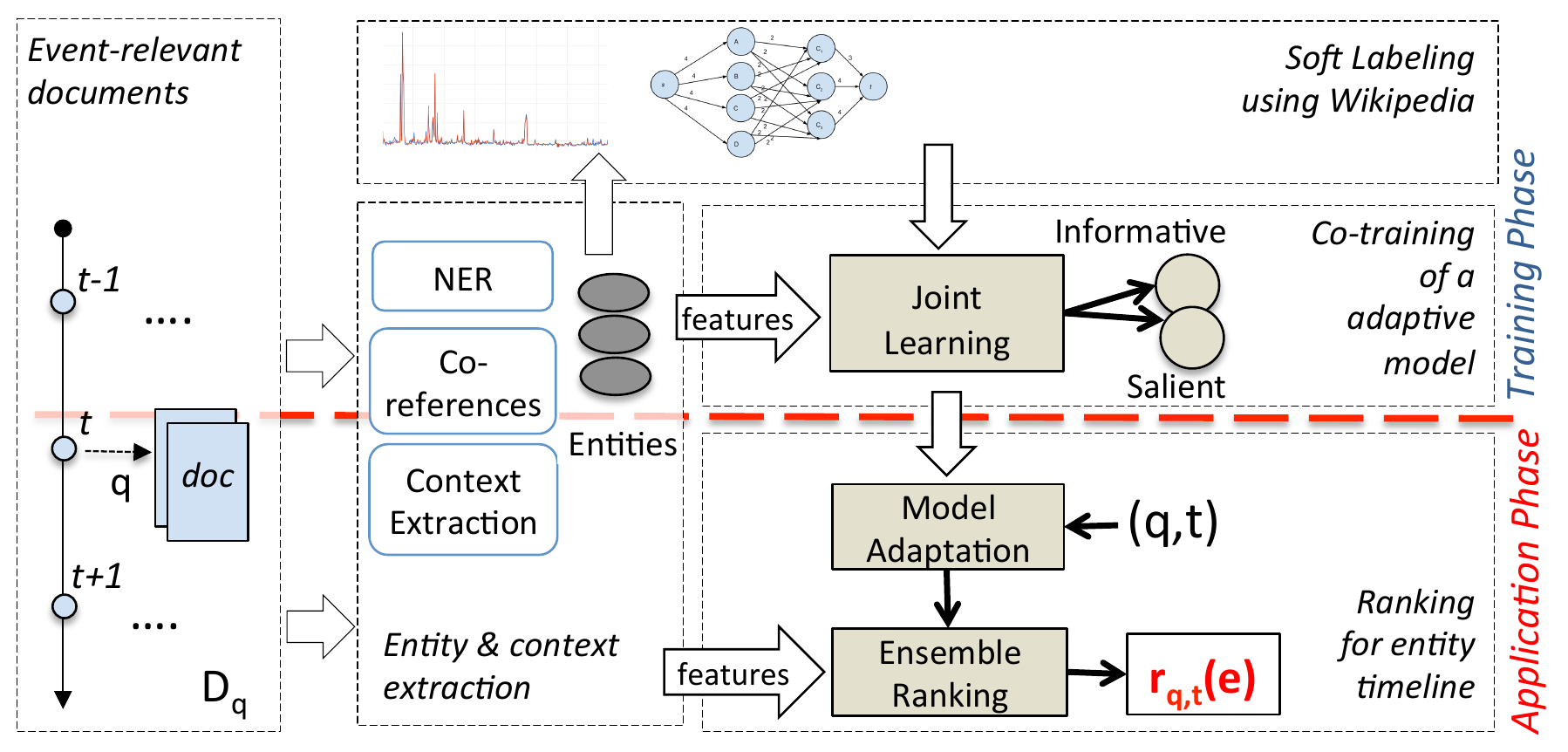}
\caption{Overview of the Entity-centric Summarization Framework}
\label{fig:arch}
\end{figure}

Figure \ref{fig:arch} gives an overview of our entity ranking framework covering both 
the training and the application/testing phase. 

Given one event $q$, its reporting timeline, and the set of documents $D_q$ (in practice, $D_q$ can be given a priori, or can be retrieved using different retrieval models) we identify the entity set $E_q$ using our entity extraction, which consists of named entity recognition, co-reference and context extraction (Section~\ref{sec:entity-extraction}). 

When the event is used for training (training phase), we link a subset of $E_q$ to Wikipedia concepts, which comprises the popular and emerging entities of the event. To facilitate the learning process, these entities are softly labeled using view statistics from Wikipedia (Section~\ref{sec:soft-labelling}), serving as training instances. Although we use popular entities for training, we design the features such that it can be generalized to arbitrary entities, independent from Wikipedia.
 
The next component in our framework is the adaptive learning that jointly learns the salience and informativeness models, taking into account the diverse nature of events and their evolution. 
(Section~\ref{sec:multi-crit-learning}).

In the application phase, entity and feature extraction are applied the same as in training phase. First, the input event and time interval is examined against the joint models to return the adaptive scores (details in Section~\ref{sec:adaptive}). Then, entities are put into an ensemble ranking, using the adapted models, to produce the final ranks for the summary.

\section{Methodology}

\subsection{Entity Extraction}
\label{sec:entity-extraction}

First, we discuss how we can extract entities, the key units in our framework. 
For each document in the set,
, we employ a named entity recognizer to identify mentions of entities in  three categories (persons, locations and organizations). For each mention, we include the containing sentences as the context.
Mentions that do not contain any alphabetical characters or only stop words are removed. 

We additionally use intra- and cross-document co-reference to track mentions pertaining to the same entity. First, an intra-document co-reference system is employed to identify all co-reference chains for entity mentions within a document. 
We include each reference in the chain together with its sentence to the set of mention contexts of the entity.
Second, to identify mentions that potentially refer to the same real-world entity across documents, we adapt the state-of-the-art cross-document co-reference method proposed by Lee et al.\cite{lee2012joint}. This method first clusters documents based on their content using an Expectation Maximization (EM) algorithm, then iteratively merges references (verbal and nominal) to entities and events in each cluster.

\textbf{Speeding up co-reference resolution:} To speed up the computation, we do not use EM clustering as \cite{lee2012joint}, but employ a set of heuristics, which have proven to be effective in practice. First, we only consider cross-document co-references from documents of the same day. Second, instead of clustering an entire document set, we use mentions with their contextual sentences (kept in the order of their appearance in the original documents) as ``pseudo-documents'' for the clustering. Third, we assume that mentions to the same entity have similar labels. Hence, we represent entity mention labels as vectors using two-grams (for instance, ``Obama'' becomes ``ob'', ``ba'', ``am'', ``ma'') and apply LSH clustering~\cite{gionis1999similarity}  
to group similar mentions. LSH has been proven to perform well in entity disambiguation tasks~\cite{hoffart2012kore}, and it is much faster than the standard EM-based clustering. 

Finally, for each entity mention, we merge its contextual sentences with those of all other references of the same co-reference chain to obtain the event-level context for an entity
, which will be used as inputs for constructing the entity features.
We note that while we are aware of other methods to increase the quality of entity extraction by linking them to a knowledge base such as YAGO or DBpedia, we choose not to limit our entities to such knowledge bases, so as to be able to identify and rank more entities in the ``long tail''. 

\subsection{Mining Wikipedia for Soft Labeling}
\label{sec:soft-labelling}

In the following, we explain how the training data is generated. Specifically, given an event $q$, an interval $t$ and an entity $e$, we aim to automatically generate the score $y^{(q)}_{j}$ such that ${y}^{(q)}_{j}(e) > {y}^{(q)}_{j}(e')$ if the entity $e$ is more prominent than the entity $e'$ with respect to the event $q$ at time $t$. This score can be used as a soft labelling to learn the ranking functions mentioned in Equation \ref{eq:l2r}.
The use of soft labeling for entities' salience has already been proposed in \cite{gamon2013identifying}, where user click behaviour in query logs is used as an indicator for entity salience scores. 
Dunnietz et al. \cite{dunietz2014new} proposed treating entities in news headlines as salient, and propagate those salience scores to other entities via the PageRank algorithm. The limitation of these measures is that they restrict the assessment of salience to the scope of individual documents, and do not consider the temporal dimension. In contrast, our soft labels are evolving, i.e., an entity can have different labels for one event at different time intervals.

The soft labeling is based on the assumption that for globally trending news event, prominence of related entities can be observed by the collective attention paid to resources representing the entities. 
For instance, during the Boston marathon bombing, Wikipedia pages about the bomber \textit{Tsarnaev} were created and viewed 15,000 times after one day, indicating their strong salience driven by the event. 
For soft labeling, we exploit the page view statistic of Wikipedia articles, which 
reflects the interest of users in an entity: Most obviously, Wikipedia articles are viewed for currently 
popular entities indicating entity salience. However, taking the encyclopedic character of Wikipedia into account, Wikipedia articles are also viewed in expectation of new information about an entity indicating its (expected) informativeness, especially in the context of an ongoing event. Actually, Wikipedia has gained attention in recent years as a source of temporal event information~\cite{Whiting:2014:WTM:2567948.2579048}. 
Event-triggered bursts in page views, as they are for example used in~\cite{Ciglan:2010:WPE:1871437.1871769}
for event detection, are thus a good proxy for the interestingness
of an event-related entity at a considered point in time, which is influenced both 
by the salience and the informativeness of the event. 

Therefore, we propose a new metric called \textit{View Outlier Ratio} or \textsf{VOR} to approximate the soft labels for a combined measure of entity salience and informativeness as follows.
For each entity $e$ and for a given time interval $t_i$, we first construct the time series of view count from the corresponding Wikipedia page of $e$ in the window of size $w$: \[T_e=[v_{e,i-w},v_{e,i-w+1},\ldots v_{e,i},v_{e,i+1},\ldots v_{i+w}]\] where each $v_{e,j}$ is the view count of the Wikipedia page of $e$ at $t_j$. From $T_e$ we calculate the median $m_{e,i}$ and define \textsf{VOR} as follows.

\begin{definition} The View Outlier Ratio is the ratio of difference between the entity view and the median:
\begin{equation}
\label{eq:pv}
vor(e_i)=\frac{|v_{e,i}-m_{e,i}|}{\mathrm{max}(m_{e,i},m_{min})}
\end{equation}
where $m_{min}$ is a minimum threshold to regularize entities with too low view activity. 

\end{definition}

\subsection{Unified Ranking Function}
\label{sec:adaptive}
We now turn our attention to defining the ranking function in Equation \ref{eq:problem}. The intuition 
is that for each event $q$ and time $t_i$, we rank an entity $e$ higher than others if: (1) $e$ is more relevant to the central parts of documents in $D_{q,i}$ where it appears (salience); and (2) the context of $e$ is more diverse to other contexts (of $e$ or other entities) at $D_{q,i-1}$. Moreover, these two criteria should be unified in an adaptive way that depends on the query, that is, event and time. For example, users interested in a festival might wish to know more about salient entities of the event, while those that follow a breaking story prefer entities with more fresh information. Even for one event, the importance of salience and informativeness might vary over time. For instance, informativeness is more important at the beginning when the event is updated frequently.
Based on this intuition, we propose the following ranking function:
\begin{equation}
\label{eq:rankfunc}
\mathbf{y_t^{(q)}} = S(q,t)\,\omega_{\mathrm{s}}^{T}\mathbf{E}_\mathrm{s} \,\,+\,\, \gamma(t)\,I(q,t)\omega_{\mathrm{i}}^{T}\mathbf{E}_\mathrm{i}
\end{equation}

\noindent where $\mathbf{E}_s$ is the $|E_q|\times M$ matrix representing the $M$ dimensional feature vectors of entities used to learn the salience score ($M$ is the number of salience features), and $\mathbf{E}_i$ is the $|E|\times N$ is the matrix of $N$ dimensional informativeness feature vectors ($N$ is the number of informativeness features). $S(q,t)$ and $I(q,t)$ represent the scores of salience and informativeness tendency for an event $q$ at $t$. Here we introduce another factor, $\gamma(t)$, which is the decay function of time $t$, controlling how much the informativeness should have impact on the overall ranking. The rationale of $\gamma$ is that when the distance between two time intervals $t_i$ and $t_{i-1}$\footnote{note that news events are not always reported on consecutive time intervals} is long, informativeness has less impact on the overall ranking. For example, if there are only reports about the news after one year (anniversary of a past event), the changes of entities in that long time period should not contribute much to the informativeness criterion.

\subsection{Multi-criteria Learning Model}
\label{sec:multi-crit-learning}

We now discuss how to learn the above ranking function using Equation \ref{eq:l2r}. A straightforward way is to learn the two models $\omega_{\mathrm{s}}$ and $\omega_{\mathrm{i}}$ separately, and assign values to $S,I$ in a pre-defined manner, then aggregate into Equation \ref{eq:rankfunc}. However, this is not desirable, since it requires to build two sets of training data for salience and informativeness at the same time, which is expensive. Secondly, previous work has pointed out that a ``hard'' classification of query based on intent itself is a difficult problem, and can harm the ranking performance \cite{geng2008query}. In this work, we exploit the \emph{divide and conquer} (DAC) learning framework in \cite{bian2010ranking} as follows. We define $\mathbf{E^{*}}$ as the $|E|\times (M+N)$ matrix of $(M+N)$ dimensional \emph{extension vectors} from the corresponding vectors of $\mathbf{E}_\mathrm{s}$ and $\mathbf{E}_\mathrm{i}$ matrices. Similarly, we define $\omega^{*}_{\mathrm{s}}$ as the $(M+N)$ extension vectors of zero vector $\mathbf{0}$, and the vectors $\omega_{\mathrm{s}}$, and $\omega^{*}_{\mathrm{i}}$ as the $(N+M)$ extension vector of $\omega_{\mathrm{i}}$ and $\mathbf{0}$. With this transformation, Equation \ref{eq:rankfunc} can be written as:

\begin{small}
\begin{equation}
\label{eq:rankfunc1}
y_t^{(q)} = S(q,t)\,g(\mathbf{E^{*}},\omega^{*}_{\mathrm{s}}) \,\,+\,\, \gamma(t)\,\,I(q,t)g(\mathbf{E^{*}},\omega^{*}_{\mathrm{i}})
\end{equation}
\end{small}

\noindent where $g(\mathbf{E^{*}},\omega)=\omega^{T}\mathbf{E^{*}}$ is a linear function. Incorporating Equations \ref{eq:l2r} and \ref{eq:rankfunc1}, we can co-learn the models $\omega^{*}_{\mathrm{s}}, \omega^{*}_{\mathrm{i}}$ (and thus $\omega_{\mathrm{s}},\omega_{\mathrm{s}}$) simultaneously, using any loss functions. 
For instance, if we use hinge loss as in~\cite{bian2010ranking}, we can then adapt the RankSVM~\cite{joachims2002optimizing}, an algorithm that seeks to learn the linear function $g$ in (\ref{eq:rankfunc1}) by minimizing the number of misordered document pairs. For completeness, we describe here the traditional objective function of RankSVM:

\begin{small}
\vspace*{-2em}
\begin{eqnarray}
\begin{aligned}
\label{eq:ranksvm}
\min_{\omega,\xi_{q,t,a,b}}\frac{1}{2}{\|\omega\|}^2 + c\sum_{q,t,a,b}{\xi_{q,t,a,b}}\ \text{\textit{,  s.t.}} & \\
\omega^{T}\mathbf{E^{*}}^{(q)}_{t,a}\geq \omega^{T}\mathbf{E^{*}}^{(q)}_{t,b} + 1 - \xi_{q,t,a,b}\forall \mathbf{E^{*}}^{(q)}_{t,a}\succ \mathbf{E^{*}}^{(q)}_{t,b}, \xi_{q,t,a,b} \geq 0
\end{aligned}
\end{eqnarray}
\end{small}

\noindent where $\mathbf{E^{*}}^{(q)}_{t,a}\succ \mathbf{E^{*}}^{(q)}_{t,b}$ implies that entity $a$ is ranked higher than entity $b$ for the event $q$ at time $t$, $\xi_{q,t,a,b}$ denotes slack variables, and $c$ sets the trade-off between the training error and model complexity. If we change the linear function $g$ to $f$, we can adapt (\ref{eq:l2r}) into (\ref{eq:ranksvm}) to obtain the following objective function:

\begin{small}
\vspace*{-2em}
\begin{eqnarray}
\begin{aligned}
\label{eq:topicranksvm}
\min_{\omega,\xi_{q,t,a,b}}\frac{\|\omega_{\mathrm{s}}\|^2+\|\omega_{\mathrm{i}}\|^2}{2} + c\sum_{q,t,a,b}{\xi_{q,t,a,b}} & & \text{s.t.}   \\
\ S(q,t)g(\mathbf{E^{*}}^{(q)}_{t,a},\omega^{*}_{\mathrm{s}}) + \gamma(t)I(q,t)g(\mathbf{E^{*}}^{(q)}_{t,a},\omega^{*}_{\mathrm{i}})\geq & \\
S(q,t)g(\mathbf{E^{*}}^{(q)}_{t,b},\omega^{*}_{\mathrm{s}}) + \gamma(t)I(q,t)g(\mathbf{E^{*}}^{(q)}_{t,b},\omega^{*}_{\mathrm{i}}) + 1 - \xi_{q,t,a,b}, & \\
\forall \mathbf{E^{*}}^{(q)}_{t,a}\succ \mathbf{E^{*}}^{(q)}_{t,b}, \xi_{q,t,a,b} \geq 0
\end{aligned}
\end{eqnarray}
\end{small}

\subsection{Event-based Models Adaptation}
\label{sec:adaptive}

The adaptive scores $S(q,t), I(q,t)$ and the decay function $\gamma(t)$ is critical to adapting the salience and informativeness models.
A na\"{i}ve supervised approach to pre-define the categories for event ($S,I$) is impractical and detrimental to ranking performance if the training data is biased. Instead, previous work on query-dependent ranking~\cite{geng2008query,bian2010ranking} often exploit the ``locality property'' of query spaces, i.e., features of queries of the same category are more similar than those of different categories. Bian et al.\cite{bian2010ranking} constructed query features using top-retrieved documents, and clustered them via a mixture model. However, the feature setting is the same for all clusters, making it hard to infer the semantics of the query categories.

In this work, we inherit and adjust the approach in~\cite{bian2010ranking} as follows. For each event $q$ and time $t$, 
we obtain all entities appearing in $D_{q,t}$ to build the ``pseudo-feedback'' for the query $(q,t)$.
We then build the query features from the pseudo-feedback as follows. From each matrix $\mathbf{E}_\mathrm{s}, \mathbf{E}_\mathrm{i}$, we take the 
\emph{mean} and \emph{variance} of the feature values of all entities in the pseudo feedback. As a result, each pair $(q,t)$ is mapped into two feature vectors (with $2M$ and $2N$ dimensions) corresponding to the salience and informativeness spaces. In each space, we use Gaussian Mixture model to calculate the centroid of the training queries, and use the distance of the query feature vector to the centroid as its corresponding adaptive score:

\begin{small}
\vspace*{-1em}
\begin{equation}
\begin{aligned}
C(q,t) = 1 - \frac{{\|\mathbf{x}^{q,t}_C-\mathbf{x}^C\|}^2}{\max_{q'\in\mathcal{Q},t'\in T_{q'}}{\|\mathbf{x}^{q',t'}_C-\mathbf{x}^C\|^2}}
\end{aligned}
\end{equation}
\end{small}

\noindent where $C\in \{I,S\}$ indicates the event categories, $\mathbf{x}^{q,t}_C$ is the query feature in the feature space of $C$, and $\mathbf{x}^C$ is the centroid of feature vectors in training set $\mathcal{Q}$ in the corresponding space. The scores are scaled between $0$ and $1$. 

\vpar
\para{Decayed Informativeness} The decay function $\gamma(t_i)$ adjusts the contribution of informativeness into the adaptive model and is defined by:

\begin{small}
\vspace*{-1em}
\begin{equation}
\begin{aligned}
\label{eq:gamma}
\gamma(t_i) = \alpha^{\lambda\frac{|t_i - t_{i-1}|}{\mu}}
\end{aligned}
\end{equation}
\end{small}

\noindent where $\lambda,\alpha$ are parameters ($0<\alpha<1,\lambda>0$), and $\mu$ is the interval unit distance. Equation~\ref{eq:gamma} represents the time impact onto the informativeness of entities: When the time lag between two intervals is high, the difference in contexts of entities between them is less likely to correlate with the informativeness quality of entities. 

\section{Entity Ranking Features}
\label{sec:features}
 We now discuss the salience and informativeness features for entity ranking. Ranking features are extracted from event documents where the entity appears as follows. These features, called \emph{individual features}, are extracted two different levels. First, 
at \textbf{context} level, features are extracted independently from each mention and its contexts. Features of this level include mention word offset, context length, or importance scores of the context within the document using summarization algorithms (SumBasic or SumFocus features). Second, at \textbf{label} level, features are extracted from all mentions, for instance aggregated term (document) frequencies of mentions. 

Based on the individual features, the entity features are constructed as follows. For each entity and feature dimension, we have the list of feature values $(z_1,z_2, \ldots, z_n)$, where $z_i$ is the individual feature of label or mention categories. For label level, we simply take the average of $z_i$'s over all entity labels. For mention level, each $z_i$ is weighted by the the confidence score of the document containing the corresponding mention and context. Such confidence score can be calculated by several ways, for instance by a reference model (e.g., BM25) when retrieving the document, or by calculating the authority score of the document in the collection (e.g., using PageRank algorithm). For all features, we apply quantile normalization, such that all individual features (and thus entity aggregated features) are scaled between $[0,1]$. Below we describe the most important features. 

\begin{table*}[hbt]
{\small
\hfill{}
\begin{tabular}{l p{5.6cm} | l p{6cm}}
\hline
\hline 
 \multicolumn{2}{c|}{\textit{Salience features}} & \multicolumn{2}{c}{\textit{Informativeness features}}\\
\hline
\textbf{Feature(s)} & \textbf{Description} & \textbf{Feature(s)} & \textbf{Description} \\
\hline
Tf / Df (M) & Term / Doc. frequency of mention in $D_{q,i}$ & PTf / Pdf (M) & Term / Doc. frequency of mention in $D_{q,i-1}$ \\
WO / SO (C) & Word / sent. offset of mention in context & CoEntE (M) & Number of co-occurring entities in $D_{q,i-1}$ \\
SentLen (C) & Context length with/without stopwords & \multirow{2}{*}{CTI(C)} & \multirow{2}{*}{\parbox{6cm}{CTI score of mention given its context in $D_{q,i}$ and $D_{q,i-1}$ \cite{wu2013measuring}}} \\
Sent-5 /-10 (C) & Context length with/without stopwords > 5/10 ? & & \\
1- / 2- / 3-Sent (C) & Is context among 1/3/5 first sentences ? & TDiv (C,M) & Topic diversity of context in $D_{q,i},D_{q,i-1}$ \\
TITLE (M) & Is the mention in titles of any $d\in D_{q,i}$ ? & CosSim (C) & Cosine similarity of context in $D_{q,i},D_{q,i-1}$ \\
CoEntM (M) & No. of entities in same context as mention & \multirow{2}{*}{DisSim (C)} & \multirow{2}{*}{\parbox{6cm}{Distributional similarity of context in $D_{q,i},D_{q,i-1}$}} \\
Sum-B / -F (C) & SumBasic / SumFocus score of context & & \\
Uni / Bi / Bi4 (C) & Uni-/Bi-/skip-Bigram overlap with query & EntDif (M) & Entity difference of context in $D_{q,i},D_{q,i-1}$ \\
Att / Sent (C) & Attitude / Sentimentality score of context & TITLEP (M) & Is the mention in titles of any $d\in D_{q,i-1}$ ? \\
Read-1 / -2 / -3 (C) & Fleisch / Fog / Kincaid readability scores 
& NewsFracP (M) & Frac. of news-specific terms
in $D_{q,i-1}$ \\
PR (C) & Sentence centrality score (PageRank) & & \\
POS (C) & POS-tag of mention in context & & \\
NewsFrac (M) & Fraction of news-specific terms co-occurring with mention & & \\
\hline
\hline
\end{tabular}}
\hfill{}
\caption{Selected Informativeness and Salience Features for Entity Ranking at Label (M) and Context (C) Level}
\label{tb:features}
\end{table*}

\subsection{Salience Features}

\para{Context importance features} One important evidence of entity salience is context of the entity mentions. It is well-known that text at the beginning of document contains more salient information~\cite{gamon2013identifying,dunietz2014new}. Besides position, the content of sentences, per se or in relations with other sentences, also indicates the salience of entities. We apply SumBasic \cite{nenkova2004evaluating} and LexRank~\cite{erkan2004lexrank} summarization algorithms to obtain the scores of contexts (features Sum-B and PR, respectively).

\vpar
\para{Human-perceived salience features} Entity salience can be assessed by reader's intentions or interests~\cite{gamon2013identifying}, and recent study suggests that user interest in entities can be attracted via serendipity in texts~\cite{bordino2013penguins}. We follow this direction and apply features presented in\cite{bordino2013penguins}, namely setimentality, attitudes, and readability scores of each mention context. For readability, we also include two other standard metrics, Gunning Fog index~\cite{fog52} and Fleisch-Kincaid index~\cite{kincaid1975derivation}.

\vpar
\para{Query-related features} Another class of salience features involves ones that are dependent on the event queries. Following~\cite{wang2013sentence}, we use the overlap between contexts and the event query at the unigram, bigram levels, and at bigram where tokens are tolerable to have 4 words in between (Bi4 feature). We also compute the query-focused SumFocus \cite{vanderwende2007beyond} score of contexts as another feature. It is worth noting that for all of these features, the event queries used are the ones that have been extended using the query expansion (see Event Document and Timelines, Section~\ref{sec:experiment}).

\subsection{Informativeness Features}

Informativeness for words has been studied extensively in linguistics community.
We adapt the state-of-the-art measure~\cite{wu2013measuring}, which incorporates the context of mentions. Given a mention $m$ of entity $e$, the context-aware informativeness score of $m$ for event $q$ at $t_i$ is defined by:

\begin{small}
\[ 
inf(m,i) = \left \{
\begin{array}{l l} 1 & U_{q,i-1} = \emptyset \\
 \frac{\sum_{c'\in U_{q,i-1}}\kappa(c',c(m))s(d_{c'},q,i-1)}{|U_{q,i-1}|}  & U_{q,i-1} \neq \emptyset
 \end{array} \right. 
 \]
\end{small}

\noindent where $U_{q,i-1}$ is the set of mention contexts of $e$ in $D_{q,i-1}$, $d_{c'}$ is the document consisting the context $c'$, $\kappa(c',c(m))$ is the sentence dis-similarity, and $s(d_{c'},q,i-1)$ is the retrieval score of $d_{c'}$. 
We normalize the metric to $[0,1]$, and set it to $1$, when the entity first appears in $D_{q,i}$. To measure $\kappa(c',c(m))$, we can employ different strategies, leading to different features:

\vpar
\parai{CTI} The most straightforward strategy is to employ a lexicon for the sentence similarity. We use the NESim method~\cite{do2009robust} for this strategy.

\vpar
\parai{Topic Diversity} Besides lexicon-based, we also calculate the informativeness on a higher level by representing contexts by latent topics, using latent Dirichlet Allocation model~\cite{Blei:2003:LDA}. Topic diversity of a context $c$ w.r.t other context $c'$ of the same entity on previous time interval is defined as
\begin{small}
\[
\kappa(c',c(m)) = \sqrt{\sum_{k=1}^{\tau}(p(\psi_k \vert c(m)) - p(\psi_k \vert c'))^2}
\]
\end{small}

\noindent where $\tau$ is the number of topics and $\psi_k$ is the topic index.

\vpar
\parai{Distributional Similarity} Another strategy uses Kullback-Leibler divergence for the dissimilarity:
\begin{small}
\begin{eqnarray}
    DisSim(c',c(m)) = -KL(\theta_{c(m)},\theta_{c'}) = \\
    -\sum_{w_i} P(w_i \vert \theta_{c(m)}) \log\dfrac{P(w_i \vert \theta_c'))}{P(w_i \vert \theta_c(m))}
\end{eqnarray}
\end{small}

\noindent where $\theta_d$ and $\theta_c$ are the language models for contexts $c',c$. We use Dirichlet smoothing method for scarce words.

\vpar
\para{Label Level} Besides context level, we also calculate a number of features at label level, by aggregating the mention features over $D_{q, i-1}$ for event $q$ at time $t_i$ (Table \ref{tb:features}). For label-level topic diversity, we use the same metric, but using concatenated contexts instead of individual ones. Additionally, we also calculate the entity difference between two context sets of the same entity in $D_{q,i-1}$ and $D_{q,i}$:
\begin{small}
\[
    EntDif(e) = \| CoEnt_i(e) \cap \overline{CoEnt_{i-1}(e)}\|
\]
\end{small}
\noindent where $CoEnt_i(e)$ is set of co-occurring entities of $e$ in $D_{q,i}$.

\section{Experiments and Evaluations}
\subsection{Setup}
\label{sec:experiment}
\para{Datasets} For our experiments, we work with a real-world, large-scale news dataset. Specifically, we use  KBA 2014 Filtered Stream Corpus (SC14) dataset, which is used for TREC 2014 Temporal Summarization track\footnote{\url{http://s3.amazonaws.com/aws-publicdatasets/trec/kba/index.html}}. We extract news and mainstream articles from the dataset, consisting of 7,592,062 documents. The dataset covers 15 long-running news events from December 2012 to April 2013. All events are high-impact, discussed largely in news media, and have their own Wikipedia pages. Each event has a predefined time span (ranging from 4 to 18 days), and is represented by one textual phrase that is used as the initial event query. Based on the event type (available in the dataset), we group the events into $4$ categories: Accident, riot and protest, natural disaster, crime (shooting, bombing). 

\para{Event Documents} To construct the event document set for the study, we firstly group the documents into individual days by their publication timestamps, and index documents for each day. 
In total, this results in 126 different indices. 
For each index, we remove boilerplate texts from document using Boilerpipe~\cite{kohlschutter2010boilerplate}, skip stop words, and lemmatize the terms. Then we use the pre-defined textual phrases of the events, issue it as a query to the corresponding indices (indices of days within the event period), retrieving from each the top 10 documents using BM25 weighting model. We improve the results using Kullback-Leibler query expansion~\cite{imran2010improving}, and add top 30 expanded terms to construct the event query $q$ used for query-related features computation. 

\para{Timelines} To build the reporting timeline $T_q$, for each event, we manually go through all the days of the event period, check the content of the top-retrieved document, and remove the day from the timeline if this top-ranked documents is not about the event. In total, we have $153$ pairs (\textit{event},\textit{day}) for all event reporting timelines. 

\para{Entities} We use Stanford parser to recognize named entities from the boilerplated content of the documents, and match them with the entities detected by BBN's Serif tool (provided in SC14 corpus) to reduce noise. For the matching entities, we use the in-document co-reference chains, which are available in SC14, and apply the cross-document coreference (Section~\ref{sec:entity-extraction}) to group mentions to entities. We use the sentences as the mention contexts. In total, we detect 72,267 named entities from the corpus, with an average of 5.04 contexts per each.

\para{Training Data} From $153$ (\textit{event},\textit{day}) pairs, we randomly choose $4$ events belonging to $4$ different categories mentioned above as a training data, resulting in $39$ pairs. To build training entities (i.e. to identify subset of entities to Wikipedia concepts, see Section~\ref{sec:entity-extraction}), we apply two named entity disambiguation softwares, WikipediaMiner and Tagme. These are the supervised machine learning tools to identify named entities from natural language texts and link them to Wikipedia. We train the models of both the tools from a Wikipedia dump downloaded in 2014 July, so as to cover all possible entities in the SC14 corpus. We only use entities co-detected by both the tools, resulting in 402 distinct entities and 665 training tuples (\textit{entity},\textit{event},\textit{day}). We use the Wikipedia page view dataset, which is publicly available, to build the soft labels for these entities. 

\para{Parameter Settings} We modify RankSVM for our joint learning tasks. Features are normalized using the Standard scaling. We tune parameters via grid search with $5$-fold cross validation and set trade-off parameter $c=20$. WikipediaMiner and Tagme tools are used with default parameter settings. For the decay parameters, given the rather small time range of events in our dataset, we empirically set $\lambda=2,\alpha=0.5$, $\mu=1$day, and leave more advance tuning for future work. For soft labeling, we set the window size $w$ to $10$ days, which is the average length of reporting timelines. The threshold $m_{min}$ is tuned as followed. For each training pair, a human expert knowing the $4$ training events well is presented with the entities, their mention contexts and content of the corresponding document. The expert is asked to put the labels on the entity from ``falsely detected'', ``non-salient entity'', ``salient but not informative'' to ``salient and informative''. Based on this judgement, we compute \textsf{VOR} scores with $m_{min}$ from $1$ to $100$, and optimize the rank of entities based on VOR scores using NDCG metric. We find that $m_{min}=12$ yields the best performance. 

\subsection{Evaluation}
\para{Baselines} We compare our approach with the following competitive baselines.

\textit{TAER}: Dermatini et al. \cite{demartini2010taer} proposed a learning framework to retrieve the most salient entities from the news, taking into consideration information from documents previously published. This approach can be considered as ``salience-pro'', since the entity salience is measured within a document, although it implicitly complements the informativeness via history documents. We train the model on the same annotated data provided by the author.

\textit{IUS}\cite{McCreadie:2014:IUS:2661829.2661951}: This work represents the ``informativeness-pro'' approach, it attempts to build update summaries for events by incrementally selecting sentences, maximizing the gain and coverage with respect to summaries on previous days. Since we are not interested in adaptively determining the cutoff values for the summary, we implement only the learning-to-rank method reported in \cite{McCreadie:2014:IUS:2661829.2661951} to score the sentences. We adapt $IUS$ into entity summarization by extracting named entities from each sentence. Then, the ranking score of the entity is calculated as the average of scores of all of its sentences across all documents.

In addition, we evaluate three other variants of our approach. The first two variants involve only salience and informativeness features for learning. We denote these as \textit{SAL} and \textit{INF}. The third variant linearly combines all salience and informativeness features, denoted as \textit{No-Adapt}. All are trained and predicted using the traditional RankSVM.

\para{Evaluation Metrics} We consider the traditional information retrieval performance metrics: precisions, NDCG and MAP. Besides, we also aim to evaluate the performance of ranking in timeline summarization context, where effective systems do not just introduce relevant, but also novel and interesting results compared to the past. This was inspired by recent work in user experience of entity retrieval~\cite{ge2010beyond,bordino2013penguins}. One popular metric widely adopted in existing work to measure such user-perceived quality is the \textit{serendipity}, which measures degree to which results are ``not highly relevant but interesting'' to user taste \cite{andre2009x}. Traditional serendipity metric was proposed to contrast the retrieval results with some obvious baseline~\cite{bordino2013penguins}. In our case, we propose to use serendipity to measure the informativeness and salience by contrasting the results of one day to previous day of the same event:

\vspace{-0.2cm}
\begin{equation}
SRDP=\frac{\sum_{e\in UNEXP}{rel(e)}}{|UNEXP|}
\end{equation}

\noindent where $UNEXP$ is the set of entities not appearing on the previous day, and $rel$ is the human relevance judgment of the entity. The relevance part ensures the salience of the entity, while the $UNEXP$ part ensures the informativeness of the entity over the event reporting timeline.
 
\vspace{-0.2cm}
\paragraph{Assessment Setup} We exclude the 39 training pairs from the overall 153 (\textit{event},\textit{day}) pairs to obtain 114 pairs for testing. For each of these pairs, we pooled the top-$10$ entities returned by all methods. In total, this results in 3,336 tuples (\textit{entity}, \textit{event}, \textit{day}) to be assessed. To accommodate the assessment, we contextualize the tuples as follows. For each tuple, we extract one sentence containing the entity from the document with the highest retrieval score (BM25), using the event as the query and on the index corresponding to the day. If there are several sentences, we extract the longest one. Next, we describe our two assessments setup, expert-based and crowdsource-based.

\para{Expert Assessment} To evaluate the quality of the systems, we employ an expert-based evaluation as follows. $5$ volunteers who are IT experts and work on temporal and event analysis were asked to assess on one or several events of their interest. For each event, the assessors were encouraged to check the corresponding Wikipedia page beforehand to gain sufficient knowledge. Then, for each tuple, we add one more contextualizing sentence, extracted from the previous date of the event. If there is no such sentence, a ``NIL'' string will be presented. We asked the assessors to check the tuple and the two sentences, and optionally, to use search engines to look for more event information on the questioned date. Then, the assessors were asked to assess the importance of the entity with respect to the event and date, in four following scales. \textbf{1}: Entity is obviously not relevant to the event; \textbf{2}: Entity is relevant to the event, but it has no new information compared to the previous day; \textbf{3}: Entity is relevant to the event and linked to new information, but it does not play a salient role in the sentence; \textbf{4}: Entity is relevant to the event, has new information, and is salient in the presented sentence. The inter-assessor agreement score for this task is $\kappa=0.4$ under the Cohen's Kappa score.

\para{Crowdsourced Assessment} In addition, we also set up a larger-scale assessment based on crowdsourcing. We use \textsf{\url{Crowdflower.com}} platform to deploy the evaluation tasks. Each tuple presented to workers consists of date, short description of the event, entity, and the sentence. Instead of direct asking for salience and informativeness of entities to event and date 
we decide for a simpler task: Asking the workers to assess entities in two steps: (1) Assessing whether the sentence is obviously relevant to the event (workers assess on a 3-point Likert scale,from ``1-Not Relavant'' to ``3-Obviously Relevant''); and (2) assessing whether the entity is important in the sentence (by virtue of being a subject or object of the sentence, binary feedback). 

Tasks are delivered such that tuples of the same \textit{(event,day)} pair go into one Crowdflower job, thus the worker has a chance to gain knowledge about event on the day and respond faster and more reliably. We pay USD $0.03$ for each tuple. To maintain the quality, we follow state-of-the-art guidelines and recommendations, and receive $5$ independent responses for each tuple. We create a gold standard for $311$ tuples, and discard responses from workers who fail to maintain an agreement of above 70\% against the gold standard. In total, we received $20\,760$ responses, $8\,940$ from which were qualified. The inter-worker agreement was $98.67\%$ under Pairwise Percent Agreement, with average variance of $42\%$, indicating a reasonably good quality given the fairly high complexity of the task.

\subsection{Results and Discussion}

\begin{table*}[htb]
\centering
\begin{tabular}{@{}cccccccc@{}}

\toprule
Method      & P@1     & P@3     & P@10    & MAP     & SRDP@1  & SRDP@3  & SRDP@10 \\
\hline \noalign{\smallskip}
\multicolumn{8}{l}{\textit{Ranking performance from expert assessment}} \\
\hline \noalign{\smallskip}
TAER        & 0.436 & 0.315 & 0.182 & 0.109 & 0.315 & 0.210 & 0.121 \\            
IUS         & 0.395 & 0.325 & 0.236 & 0.141 & 0.335 & 0.217 & 0.176 \\                             
SAL         & 0.493{$^\blacktriangle$} & 0.423 & 0.338{$^\blacktriangle$} & 0.217{$^\blacktriangle$} & 0.421 & 0.320 & 0.240{$^\blacktriangle$} \\                             
INF         & 0.480{$^\blacktriangle$} & 0.436 & 0.354{$^\blacktriangle$} & 0.227{$^\blacktriangle$} & 0.441{$^\blacktriangle$} & 0.340 & 0.256{$^\blacktriangle$}\\    
MAX(S,I)    & 0.493 & 0.436 & 0.354 & 0.227 & 0.441 & 0.340 & 0.256\\     
\midrule
No-Adapt    & 0.503 & 0.461 & 0.320 & 0.225 & 0.396 & 0.338 & 0.215 \\                             
AdaptER     & \textbf{0.546} & \textbf{0.485} & \textbf{0.368} & \textbf{0.264} & \textbf{0.507$^\blacktriangle$} & \textbf{0.440$^\blacktriangle$} & \textbf{0.275} \\       
\hline \noalign{\smallskip}
\multicolumn{8}{l}{\textit{Ranking performance from crowdsourced assessment}} \\
\hline \noalign{\smallskip}
TAER        & 0.229 & 0.183 & 0.106 & 0.066 & 0.201 & 0.146 & 0.079 \\
IUS         & 0.258 & 0.202 & 0.154 & 0.092 & 0.197 & 0.165 & 0.119 \\
SAL         & 0.320 & 0.279 & 0.207 & 0.139 & 0.279 & 0.218 & 0.154 \\
INF         & 0.313 & 0.283 & 0.214  & 0.146 & 0.306 & 0.229{$^\blacktriangle$} & 0.160 \\
MAX(S,I)         & 0.320 & 0.283 & 0.214 & 0.146 & 0.306 & 0.229 & 0.160 \\
\midrule
No-Adapt    & 0.271 & 0.252 & 0.181 & 0.123 & 0.236 & 0.208 & 0.144 \\
AdaptER    & \textbf{0.388$^\blacktriangle$} & \textbf{0.340} & \textbf{0.237} & \textbf{0.178} & \textbf{0.361} & \textbf{0.361$^\blacktriangle$} & \textbf{0.315$^\blacktriangle$} \\
\hline \noalign{\smallskip}
\end{tabular}
\captionsetup{justification=centering,margin=1cm}
\caption{Entity-ranking performance using different assessment settings. In each setting, significance is tested against line 1, TAER (within the first group), and line 5, MAX(S,I) (within the second group). Symbol $^\blacktriangle$ indicates cases with confirmed significant increase}
\label{tbl:result}
\end{table*}

The upper part of Table \ref{tbl:result} summarizes the main results of our experiments
from the expert evaluation. The results show the 
performance of the two baselines (\textit{TAER} and \textit{IUS}) and of the 
consideration of Salience and Informativeness features in isolation 
with respect to precision. In general, all performances are low, indicating the relatively high complexity of this new task. In addition, as can be seen from this part 
of the table, even the approach relying on our salience features 
or informativeness features in isolation already outperforms the two baselines. 
This is due to the fact that our approach does not consider 
documents in isolation as the baselines do. Rather, we take 
a more comprehensive view considering event level 
instead of document level features via feature aggregation. In more details, the first baseline (\textit{TAER}) employs a quite restricted features for entity ranking (e.g. document frequency), and thus fails to identify important entities event-wise.

Furthermore, the results also show the 
performance of the non-adaptive combination of salience and 
informativeness (\textit{No-Adapt}) as well as our approach (\textit{AdaptER}), 
which uses an adaptive combination of informativeness 
and saliency.  It becomes clear that an
improvement by combining the salience and the 
informativeness features over the use of the isolated features 
can only be achieved by fusing the two features 
in a more sophisticated way: \textit{No-Adapt} does not perform better 
than the maximum of \textit{SAL} and \textit{INF} ($MAX(S,I)$), it even performs worse in 
under some metrics such as P@10. In contrast, \textit{AdaptER} clearly 
outperforms the maximum of \textit{SAL} and \textit{INF} as well as its 
non-adaptive version for most metrics. For instance, we achieve 16\% improvement of MAP scores as compared with the $MAX(S,I)$. 



Besides precision, we also consider
serendipity (SRDP) as a complementary measure in our experiments, as discussed above. This metric measures how likely the approach brings unseen and interesting results to the user. Under SRDP, our approach outperforms significantly
both the baseline and the maximum of \textit{SAL} and \textit{INF}. We achieve $14\%$ improvement of serendipity at top-$1$ entities, and $29\%$ at top-$3$ entities. Thus, our top-retrieved entities do not only cover relevance, but are also more interesting, often unseen on the previous day (contributing to more informative results).

The lower part of the Table~\ref{tbl:result} shows the same results for the crowdsourced assessment. 
The same trends of performance can be observed, where our approach outperforms in all the metrics. In comparison to the expert evaluation, the
results are overall lower. A possible reason for this is the complexity of the crowdsourcing task, which requires knowledge about the considered event in order to give high quality feedback (see Expert Assessment setup, Section~\ref{sec:experiment}). Nevertheless, the adaptive model is still able to achieve significant gain, especially under the serendipity measurement. 

\begin{table}[htb]
\centering
\footnotesize
\begin{tabular}{@{}l l l l @{}}
\hline \noalign{\smallskip}
\multicolumn{2}{l}{}                  & P@10     & MAP  \\
\hline \noalign{\smallskip}
No-Adapt        & Soft-labeled        & 0.32    & 0.225  \\
                & Supervised          & 0.346 & 0.247 \\
\hline \noalign{\smallskip}
AdaptER        & Soft-labeled        & 0.368   & 0.264 \\
                & Supervised          & 0.357   & 0.259  \\
\hline \noalign{\smallskip}
\end{tabular}
\caption{Performance of systems with(-out) soft labeling}
\label{table:softlabel}
\end{table}

Next, we evaluate the effectiveness of soft labeling in covering salience and informativeness. For this purpose, we manually annotate entities obtained from the reporting timeline of $4$ training events, with respect to the salience and informativeness (see Parameter settings, Section~\ref{sec:experiment}). We then re-train both non-adaptive and adaptive models using this annotated data (supervised approach). Table~\ref{table:softlabel} shows the precision and MAP scores of the supervised approach in comparison to our soft labelling-based approach. The similarity in performance between them, regardless of the models they have been used for, confirms that our soft labelling properly captures both salience and informativeness. 


\vpar
\para{Feature Analysis}
Analysing the influence of different feature groups (see Section~\ref{sec:features}) can give insights into what factors contribute to the entity ranking performance. To study the feature impacts, we do an ablation study and remove incrementally a group of features, and re-evaluate the performance using the expert assessment. Since reducing the feature dimensions directly affects the query features and its adaptive scores, to ensure the fair comparison, we perform the study for the non-adaptive setting. Table~\ref{table:fa} shows the MAP scores of ablated models. The symbol $\blacktriangledown$ indicates a significant decrease with respect to \textit{No-Adapt} model (with full features), and thus implies the high influence of the corresponding feature group. From the table, we can see that the most influential feature groups include context importance feature (salience features) and informativeness feature group of context level.

\begin{table}[htb]
\centering
\footnotesize
\begin{tabular}{@{}l l l @{}}
\hline \noalign{\smallskip}
Ablated feature set & MAP  \\
\hline \noalign{\smallskip}
Context importance       &  0.086$\blacktriangledown$  \\
Human-perceived & 0.097 \\
\hline \noalign{\smallskip}
Informativeness, context level & 0.06$\blacktriangledown$ \\
Informativeness, label level & 0.120 \\
\hline \noalign{\smallskip}
\end{tabular}
\caption{MAP of No-Adapt when feature groups are ablated (full-feature MAP score: 0.225)}
\label{table:fa}
\end{table}

\para{Anecdotic Example}
In table~\ref{table:WikiPortalExamples}, we show one example of top-selected entities for the event ``Boston marathon bombing 2013''. Additionally, we show some selected sentences covering the entities, to enable the understanding of the entities' roles within the event on the presented days. As can be seen, the timeline corresponding to \textit{TAER} approach (upper part) gives more salience credits to entities frequently mentioned throughout the news (such as Boston marathon), keeping them in high ranks throughout the timeline. The approach is not responsive to less salient but interesting entities (such as Pop Francis, a rather unrelated entity to the event, but get involved via his condolecence and activities to victims of the bombing). On the other hand, using an adaptive ranking with informativeness incorporated, the resulting entities are not just more diverse (including related events such as Marathon Bruins), but also expose more new and emerging information. 

\begin{table*}
\scriptsize
\centering
\begin{tabular}{ |p{1.7in}|p{1.7in}|p{1.7in}| }
  \hline
  \textit{April 15} & \textit{April 16} & \textit{ April 17} \\ \hline \hline
\textbf{Boston Marathon} \newline \textbf{Mass General Hospital} \newline \textbf{Boston.com} & \textbf{Boston} \newline \textbf{Boston Marathon} \newline \textbf{Vatican} & \textbf{Boston Marathon} \newline \textbf{Boston} \newline \textbf{Boston University} \\ \hline
- Two bombs exploded near the finish of the Boston Marathon on Monday, killing two people, injuring 22 others \newline - At least four people are in the emergency room at Mass General Hospital &  Deeply grieved by news of the loss of life and grave injuries caused by the act of violence perpetrated last evening in Boston, His Holiness Pope Francis wishes me to assure you of his sympathy \ldots  & - FBI confirmed that pressure cookers may have been used as explosive devices at the Boston Marathon. \newline - The third victim was identified Wednesday as Boston University graduate student Lingzi Lu.
\\
\hline
\hline
\textbf{Boston Marathon} \newline \textbf{Marathon Bruins} \newline \textbf{New York City} & \textbf{Pope Francis} \newline \textbf{Vatican} \newline \textbf{Boston Marathon} & \textbf{FBI} \newline \textbf{Boston University} \newline \textbf{Lingzi Lu} \\ \hline

- Two bombs exploded near the finish of the Boston Marathon on Monday, killing two people, injuring 22 others \newline - The NHL postponed the Boston Bruins' Monday hockey game due to the bombing &  The Vatican sent a telegram to Boston Cardinal on Tuesday, in which Pope Francis expresses sympathy for the victims of the marathon bombings\ldots  & - FBI confirmed that pressure cookers may have been used as explosive devices at the Boston Marathon. \newline - The third victim was identified Wednesday as Boston University graduate student Lingzi Lu.\\ \hline
\end{tabular}
\captionsetup{justification=centering,margin=2cm}
\caption{Examples of top-$3$ entities on Boston Marathon Bombing 2013 using TAER (top) and AdaptER(bottom) for April 15, 16 and 17}
  \label{table:WikiPortalExamples}
\end{table*}

\section{Conclusions}
In this paper, we have presented a novel approach for timeline summarization of high-impact news events, using entities as the main unit of summary. We propose to dynamically adapt between entity salience and informativeness for improving the user experience of the summary. Furthermore, we introduce an adaptive learning to rank framework that jointly learns the salience and informativeness features in a unified manner. To scale the learning, we exploit Wikipedia page views as an implicit signal of user interest towards entities related to high-impact events. Our experiments have shown that the introduced methods considerably improve the entity selection performance, using both small-scale expert-based and large-scale crowdsourced assessments. The evaluation also confirms that integrating salience and informativeness can significantly improve 
the user experience of finding surprisingly interesting results. 

As the problem discussed in the paper is new, there are several promising directions to explore for future work. We aim to investigate further the impact of adaptation in different types of events, using larger and more diverse sets of events. Furthermore, we plan to study more advanced ways to mine Wikipedia temporal information as signals of collective attention towards public events. We are planning to use this for further improving our VOR measure for the soft labeling approach. Other direction includes investigating deeper models to improve the performances of current entity timeline summarization systems, which are quite low.

\vspace{0.25cm}
\noindent\textbf{Acknowledgements.} {This work was partially funded by the European
Commission for the FP7 project ForgetIT (under grant No. 600826) and the ERC Advanced Grant ALEXANDRIA (under grant No. 339233).
}

\begin{small}
\bibliographystyle{abbrv}

\end{small}
\end{document}